\documentclass[aps,prl,twocolumn,superscriptaddress,groupedaddress]{revtex4}  

\usepackage{graphicx}  
\usepackage{dcolumn}   
\usepackage{bm}        
\usepackage{amssymb}   
\usepackage{overpic}
\usepackage[usenames]{color}
\usepackage{hyperref}
\usepackage{multirow}
\usepackage{ulem}
\usepackage{xc}
\usepackage{times}

\usepackage{overpic}

\usepackage{multirow}
\usepackage{ulem}

\begin{document}

\title{Metabolic plasticity in synthetic lethal mutants: viability at higher cost}

\author{Francesco Alessandro Massucci}
\affiliation{Departament de F{\'\i}sica de la Mat\`eria Condensada, Universitat de
  Barcelona, Mart\'{\i} i Franqu\`es 1, 08028 Barcelona, Spain}
  
 \author{Francesc Sagu\'es}
\affiliation{Departament de Qu{\'\i}mica F{\'\i}sica, Universitat de
  Barcelona, Mart\'{\i} i Franqu\`es 1, 08028 Barcelona, Spain}
  
 \author{M. \'Angeles Serrano}
 \affiliation{Departament de F{\'\i}sica de la Mat\`eria Condensada, Universitat de
  Barcelona, Mart\'{\i} i Franqu\`es 1, 08028 Barcelona, Spain}

 \affiliation{Universitat de Barcelona Institute of Complex Systems (UBICS), Universitat de Barcelona, Barcelona, Spain}
 
  \affiliation{ICREA, Pg. Llu\'is Companys 23, 08010 Barcelona, Spain}
 
\date{This manuscript was compiled on \today}

\begin{abstract}
The most frequent form of pairwise synthetic lethality (SL) in metabolic networks is known as plasticity synthetic lethality (PSL). It occurs when the simultaneous inhibition of paired functional and silent metabolic reactions or genes is lethal, while the default of the functional reaction or gene in the pair is backed up by the activation of the silent one. 
Based on a complex systems approach and by using computational techniques on bacterial genome-scale metabolic reconstructions, we found that the failure of the functional PSL partner triggers a critical reorganization of fluxes to ensure viability in the mutant which not only affects the SL pair but a significant fraction of other interconnected reactions forming what we call a SL cluster. Interestingly, 
SL clusters show a strong entanglement both in terms of silent coessential reactions, which band together to form backup systems,
and of other functional and silent reactions in the metabolic network. 
This strong overlap, also detected at the level of genes, mitigates the acquired vulnerabilities and increased structural and functional costs that pay for the robustness provided by essential plasticity. Finally, the participation of coessential reactions and genes in different SL clusters is very heterogeneous and those at the intersection of many SL clusters could serve as supertargets for more efficient drug action in the treatment of complex diseases and to elucidate improved strategies directed to reduce undesired resistance to chemicals in pathogens.

\end{abstract}

\maketitle

{\small \section{Significance Statement}{Synthetic lethality (SL), in which the combined knockout of two nonessential genes or reactions is lethal, has direct applications in recognising targets for therapeutic treatment of complex diseases and for fighting against undesired resistance. Typically, SL interactions are reported in pairs. We propose a change of paradigm based on the fact that SL interactions in metabolism are not independent of each other but form complex backup systems involving the rearrangement of a significant SL cluster of metabolic fluxes to ensure viability. This robustness comes at the expenses of acquired vulnerabilities and increased costs, mitigated by the entanglement of SL clusters in terms of shared reactions and genes, which could serve as supertargets for a new generation of therapeutic treatments.
}
}



\section{Introduction}\label{sec:introduction}

In metabolic networks, phenotypic responses to mutations that block the activity of nonessential biochemical reactions imply a fast rearrangement of fluxes. This metabolic plasticity, understood as a reorganisation or repair of damage in response to a disruption, is a signal of the robustness of metabolism against perturbations~\cite{Wagner:2005b}. Further indications of metabolism robustness is provided by experimental results showing that more than $90\%$ of the genes in {\it Escherichia coli} K-12 are probably not essential, with metabolic genes presenting no exception~\cite{Baba:2006}. 

However, among the viable mutations, some are critically fragile. If a mutated enzyme-coding gene or a disrupted reaction forms a synthetic lethal (SL) pair with a partner, meaning that their simultaneous deletion becomes lethal for the organism even though the individual removals are not~\cite{Hartman:2001,Tucker:2003a,Deutscher:2006,Suthers:2009,Guell:2014b}, metabolic plasticity becomes essential to ensure viability. These synthetic lethalities provide the mutant with new vulnerabilities, exploitable  for antimicrobial drug target identification~\cite{Roemer:2013} or, in the case of eukarya, for cancer therapy~\cite{Jerby:2014}. 

Due to the complex interconnectivity of metabolic networks~\cite{Jeong:2000,Ma:2003b,Guimera:2005b}, the critical reorganisation of fluxes in most SL mutants may affect a significant fraction of reactions other than the SL pair. This would be specially relevant for SL interactions formed by a functional or active (non-zero flux) and a silent or inactive (zero flux) reaction in the native state.These interactions are called plasticity synthetic lethal (PSL) pairs, the dominant SL category in {\it Escherichia coli}~\cite{Guell:2014b}. Mutants of PSL interactions, in which the inactive reaction in the pair activates as a backup when the active reaction fails, could be impaired by a metabolic burden which should be compensated by specific metabolic/genetic mechanisms. A parallel situation has been observed, for instance, in antibiotic resistant mutants. A nonspecific metabolic stress leading to a potential fitness cost~\cite{Anderson:1999} has been proposed to be compensated for by adjusting metabolism without the need for acquiring compensatory mutations~\cite{Olivares:2014}.

In the following, we use computational techniques on genome-scale metabolic reconstructions~\cite{Orth:2010a} of three bacteria in the {\it Escherichia} genus to define and characterize the metabolic reorganisation related with essential plasticity in PSL mutants. We quantify the increased structural and functional metabolic burden of essential plasticity, and we explore the mechanisms that buffer against the huge costs expected for sustaining alternative backup mechanisms for all PSL pairs in these organisms. 

\section{Results}\label{sec:results}
We studied three bacteria in the {\it Escherichia} genus: {\it Escherichia coli} K-12 MG1655 {\it i}JO1366~\cite{Orth:2011a} ({\it E. coli}), {\it Shigella sonnei} Ss046 {\it i}SSON$\_$1240~\cite{Monk:2013} ({\it S. sonnei}), and {\it Salmonella enterica} enterica serovar Typhimurium LT2 STM$\_$v1$\_$0~\cite{Thiele:2011} ({\it S. enterica}). We simulated {\it in silico} metabolic fluxes on each genome-scale reconstruction using Flux Balance Analysis (FBA)~\cite{Orth:2010} with optimization of biomass yield in glucose minimal medium, and assumed viability when the flux through the biomass function was nonzero (see Methods). By applying FBA to mutants defective in single or pairs of reactions, we identified all {\it in silico} single essential reactions and SL reaction pairs in the organisms, as in~\cite{Deutscher:2006,Suthers:2009,Guell:2014b}. We classified synthetic lethal combinations as Plasticity SL (PSL) pairs, with one reaction functional and one silent in the WT, and Redundancy SL (RSL) pairs, with both reactions active~\cite{Guell:2014b}. In each species, more than $75\%$ of the pairs are in the PSL class (see Table 1 in Methods). Next, we focus on this category both for their abundance and also for the impact that the failure of the functional PSL reaction has in terms of flux reorganization.

\subsection{Definition of SL cluster}
We name the WT functional reaction in a PSL pair a metabolic switch, and the WT silent coessential partner its backup. When the metabolic switch fails, phenotypic reorganization takes place to allow viability. More specifically, the inactive reaction in the PSL pair turns on acting as a functional backup which buffers against the mutation. Notice that one switch can have more than one backup, with proportions close to 4 backups for every switch in {\it E. coli} and {\it S. sonnei} 2 to 1 in  {\it S. enterica} (see Methods, Table 1). For the three bacteria, all the backups associated with the same switch activate together, forming what we call a backup system. 

\begin{figure}[t!]
 \centering
\includegraphics[width=8.7cm]{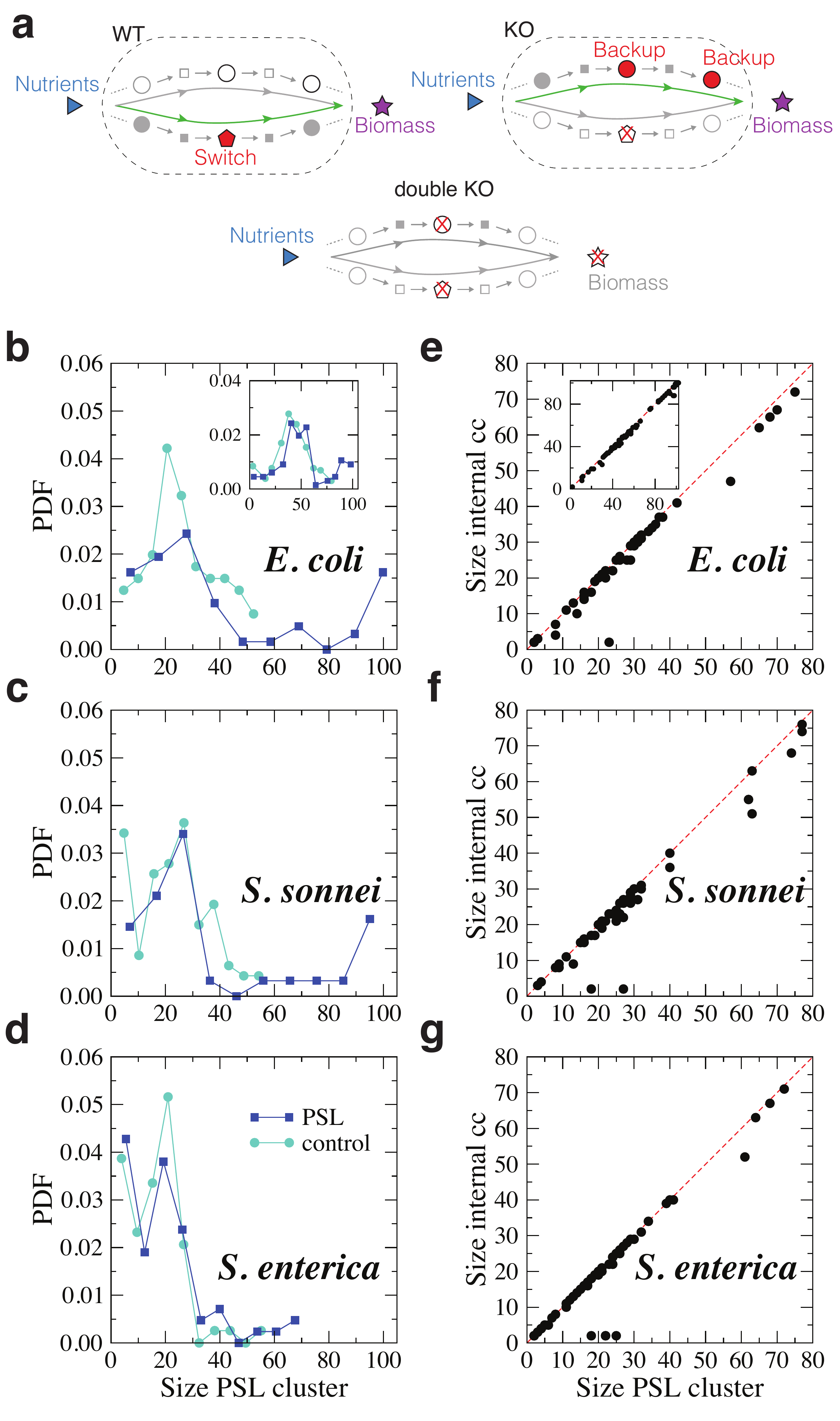}
 \caption{\textbf{Sketch of a SL cluster and size statistics.} %
 \textbf{a.} In the WT, the metabolic flow happens through a switch reaction (the red pentagon). When it fails, the metabolic flow goes through an alternative channel, that encompasses the backup system of the switch (red circles). When both the switch and one of its backups are deleted, growth may no longer be attained. The SL cluster is denoted by all reactions inside the dashed box.  \textbf{b-d.}  Probability distribution functions of SL clusters sizes for PSL mutants in minimal medium as compared to the control (given by mutants of active, nonessential and noncoessential reactions). \textbf{e-g.} Number of reactions in SL clusters as compared to the sizes of the corresponding internal connected components in minimal medium. \textbf{Inset e.} The same as in \textbf{e} for {\it E. coli} in rich medium.}
 \label{fig:1}
\end{figure}

At the same time, other fluxes reorganize such that reactions that were inactive in the WT become active in the mutant and vice versa. For every switch of {\it E. coli}, {\it S. sonnei} and {\it S. enterica}, we identified the differentially activated reactions in the mutant which changed from active to inactive or from inactive to active. The whole set is named the SL cluster, Fig.~(\ref{fig:1})a. By definition, the SL cluster contains the switch and its backup system and the switch is unique to a SL cluster and cannot enter as a backup in any other. However, backups can serve more than one switch, and both switches and backups can enter as differentially activated reactions in other SL clusters as well. In the studied organisms, the SL clusters have a small but significant size, comprising groups of about $5\%$ of the total number of reaction in the genome-scale reconstructions. The size distributions for all SL clusters in the three bacteria are shown in Fig.~(\ref{fig:1})b-d. As a control, we use the differentially activated reaction sets resulting in mutants obtained when reactions which are active in WT but not essential or coessential are knocked out. The comparison shows that the size distribution of SL clusters has, for {\it E. coli} and {\it S. sonnei}, a long tail implying that essential plasticity is more complex and involves more metabolic reactions than reorganizations induced by nonessential mutations. This is also valid for results in rich medium, see inset Fig.~(\ref{fig:1})b.

\begin{figure*}[t!]
 \centering
 \includegraphics[width=0.9\textwidth]{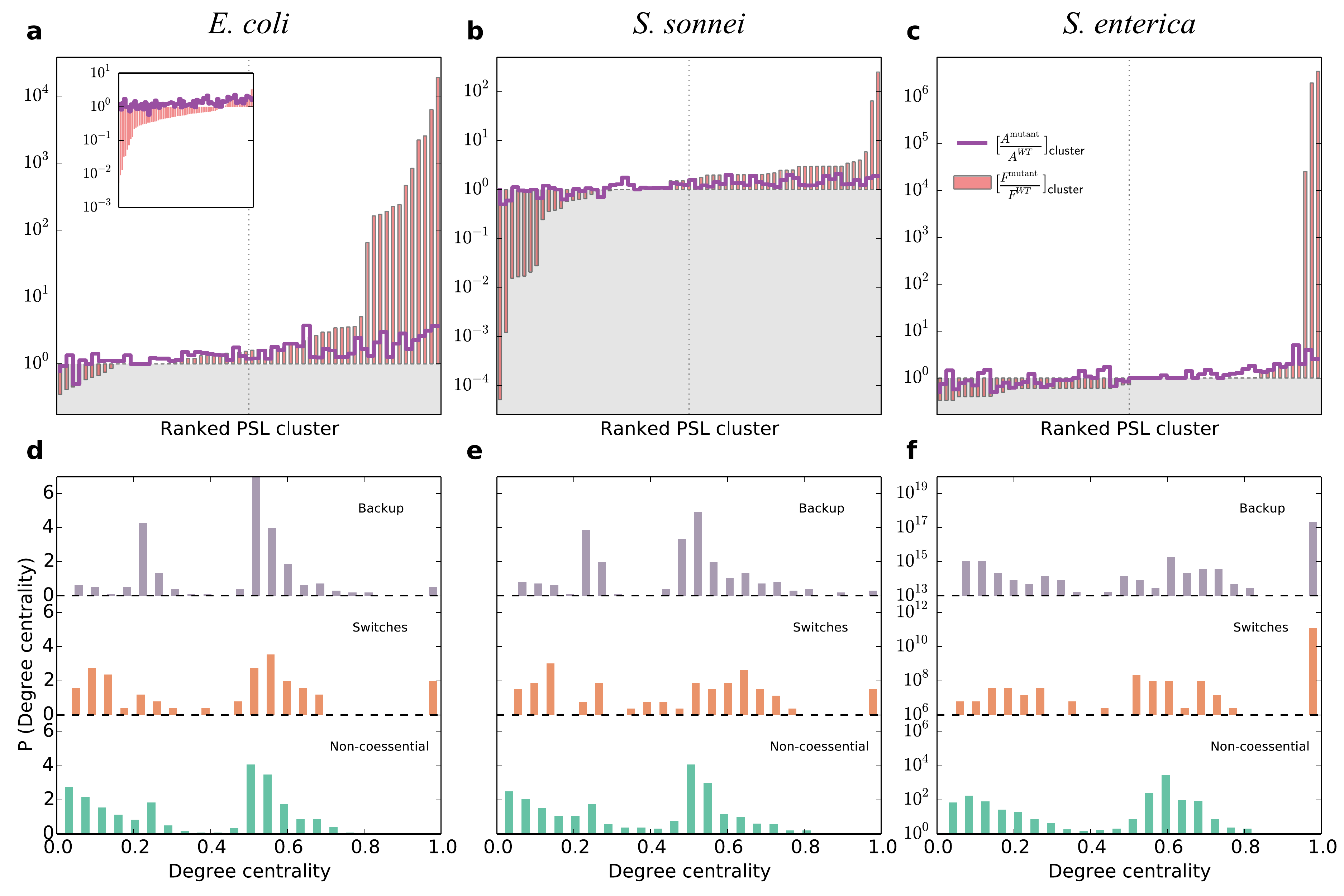}
 \caption{\textbf{Cost of essential plasticity.} \textbf{a -- c) Functional cost.} For the three bacteria in minimal medium, we show the PSL-mutant--to--WT ratio of the total flux running through SL clusters (bars), the number of active reactions in SL clusters (continuous line), and the ATP production rate obtained by optimizing growth (discontinuous line). The grey shaded area indicates the region where the PSL mutants have values smaller than in the WT. The dotted vertical lines indicate half of the number of PSL mutants. \textbf{Inset a.} The same as in \textbf{a} for rich medium. \textbf{d--f) Structural cost.} Distributions of the degree centrality of reactions belonging to the switch, backup, and noncoessential reactions inside SL clusters for the three organisms considered.}
 \label{fig:2}
\end{figure*}

Reactions in SL clusters form cohesive structures. First, the switch belongs to a connected component of differentially activated reactions which includes its backups~\footnote {The only exception corresponds to the SL cluster associated to the switch reaction phosphoenolpyruvate carboxylase in {\it S. enterica}. It has Isocitrate lyase as its only backup, which is not included in the internal connected component but forms an isolated component inside the corresponding SL cluster.}. Second, this connected component is the largest in the SL cluster and on average includes more than $92\%$ of its reactions, Fig.~(\ref{fig:1})e-g. The rest form residual disconnected components scattered through the metabolic network. Only 2 clusters in {\it E. coli} and {\it S. sonnei} and 4 in  {\it S. enterica} out of approximately 60 in each organism (see Table I in Methods) deviate from this behavior. The explanation for the divergence of the sizes of the SL cluster and its connected component is most frequently a change of strategy in mutants, like the switch from aerobic to anaerobic metabolism associated with the extracellullar transport of oxygen in {\it E. coli}, and to the reaction protoporphyrinogen oxidase in {\it S. sonnei}~\footnote {Special mention deserves the SL cluster in the three organisms associated with the extracellular transport of glucose via diffusion. Both the switch and the backup are connected to exactly the same reactants and products. Therefore, they are equivalent alternatives for FBA, meaning that we cannot distinguish between switch and backup because both reactions in the coessential pair can indistinctly play both roles. As a consequence, the corresponding internal connected component is only formed by the switch and the backup while the SL cluster contains other reactions. We found no other topologically invariant coessential PSL pairs in the organisms.}. 

\subsection{Functional and structural cost of essential plasticity}

The fact that coessential partners of switch reactions in PSL pairs remain silent in the WT and only change their activation state when needed to ensure the viability of the organism points to possible costs associated to the activation of these backup systems. To check this hypothesis, we measured both structural and functional costs associated to the viability of PSL mutants~\footnote{Notice that the biomass yield is not a good indicator of possible functional costs. In more than $80\%$ of the mutants the FBA solution is suboptimal but very close to optimal, while in the remaining growth is not changed with respect to the WT.}. 

We quantified the flux and energetic requirements of PSL mutants as compared to the WT. The flux disparity measure is given by the ratio of the total flux running through reactions in SL clusters in mutants relative to that of the same reactions in the WT, $[F^{mutant}/F^{WT}]_{cluster}$. The results in Fig.~(\ref{fig:2})a-c show that this quantity is specific to the mutant. However, in {\it E. coli} and {\it S. sonnei} more than $70\%$ of the clusters have a mutant-to-WT flux ratio larger than one with average values $542.7$ and $6.6$, respectively. Conversely, the flux through more than half of the SL clusters of {\it S. enterica} is lower in PSL mutants than in the WT, but flux decreases are relatively minor in most PSL mutants while flux increases are dramatic, such that the average of the ratio over mutants is more than $90000$ (Methods, Table I). 

The observed flux increase is not only related to the larger number of active reactions in SL clusters of mutants as compared to the WT, Fig.~(\ref{fig:2})a-c and Supplementary Fig.~(S1), but also to a higher average flux per active reaction, see Supplementary Fig.~(S2). More specifically, 53 mutants in {\it E. coli} have more active reactions in their SL cluster than in the WT, and the average flux per active reaction is higher in $69\%$ of the clusters. In {\it S. sonnei}, the number of mutants with more active reactions in the SL cluster than in the WT is still very high, 49 of 63, but the situation is more balanced regarding the average flux per active reaction. In {\it S. enterica}, still 40 of the 61 mutants have the same or more active reactions in SL clusters but the average flux per active reaction is generally lower.

The second magnitude calibrating the functional cost of viability in PSL mutants is given by the ratio of ATP production in the mutants as compared to the WT, $E^{mutant}/E^{WT}$. ATP production is defined as the flux through the ATP maintenance reaction --which is a balanced ATP hydrolysis reaction used to simulate energy demands not associated with growth-- versus the intake flux of oxygen, both obtained by optimizing biomass yield~\cite{Guell:2014b}. The flux through the ATP maintenance reaction is preserved and deviations of the ATP production in some mutants as compared to WT are basically due to changes in oxygen consumption, see Supplementary Fig.~(S3). The number of mutants showing this altered phenotype is $35\%$ in {\it E. coli} and $41\%$ in {\it S. sonnei}. Oxygen needs for the production of the same ATP amount in the affected {\it E. coli} mutants is always increased, while it is always decreased in altered {\it S. sonnei} mutants. Curiously, the variations in oxygen consumption in mutants of both bacteria is in all cases a fixed amount.

The increased number of active reactions in most mutants gives a rough indication of the cost of essential plasticity at the structural level. A different estimation can be obtained by evaluating the degree centrality~\cite{Newman:2010} of reactions within the internal connected component of a SL cluster, which informs about their role in mediating interactions among other reactions. The investigation of this quantity for the three bacteria, shown in Fig.~(\ref{fig:2})d-f (the results are qualitatively similar for other centrality measures), reveals that PSL coessential reactions, and specially backups, have higher centrality as compared to noncoessential reactions. The centrality distribution of switches features a tendency towards lower centrality values, but they always possess a marked peak at very large values. This makes them, on average, more central than noncoessential reactions. Finally, backup reactions have a distribution which is more concentrated at intermediate-to-large centrality values. In fact, the probability that backups have higher centrality than switches and than noncoessential reactions in {\it E. coli} and {\it S. sonnei} is approximately $0.64$ and $0.55$, respectively, while the value for {\it S. enterica} is notably smaller, $0.37$.

\subsection{Entanglement of SL clusters by overlapping reactions}
\begin{figure*}[t!]
\centering
 \includegraphics[width=0.9\textwidth]{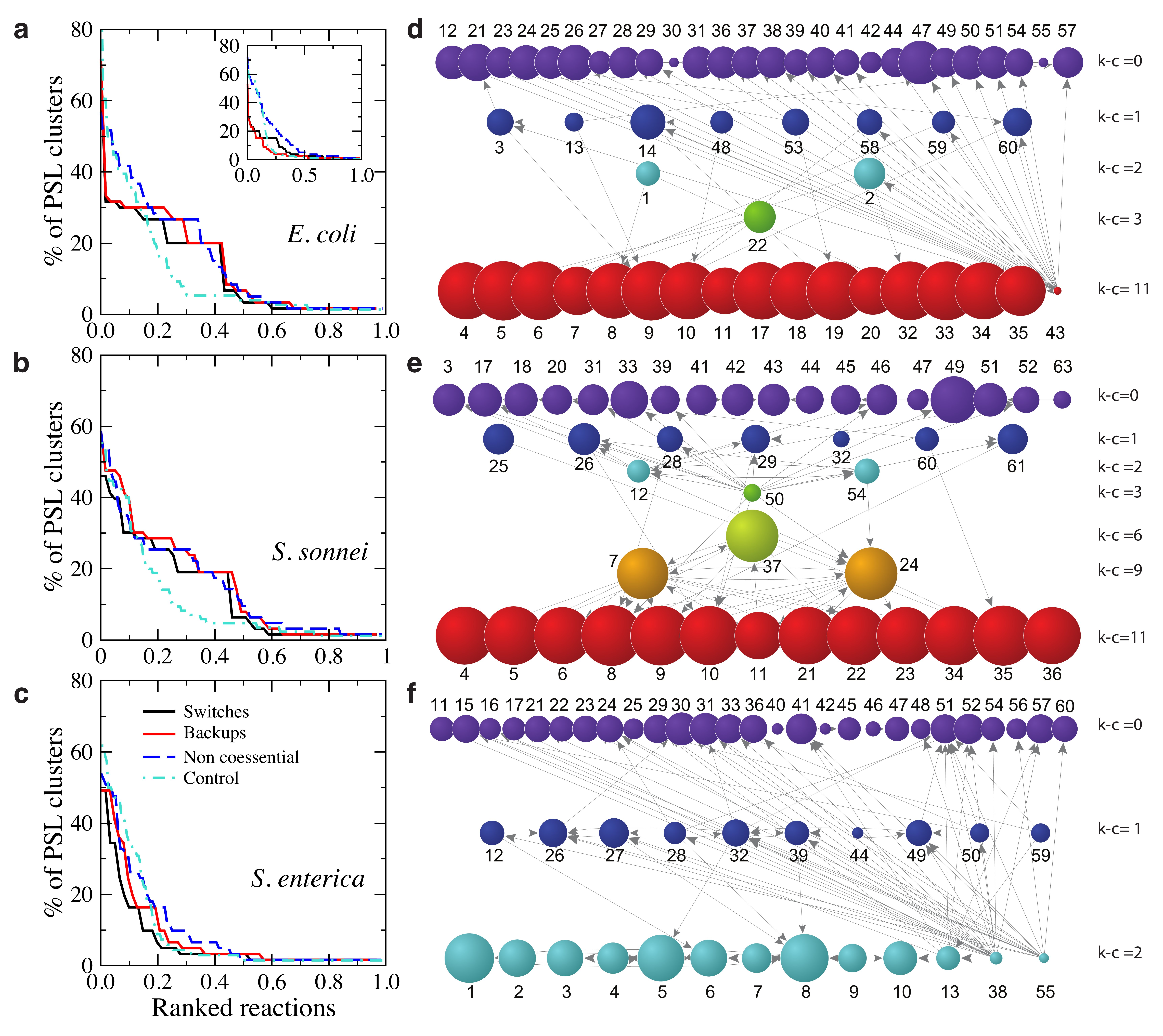}
 \caption{\textbf{Redundancy of reactions in SL clusters.} \textbf{a --c}: {\bf Ranking of reactions ordered by number of occurrences in different SL clusters in minimal medium}. Switches, backup, and non--coessential reactions in the SL clusters and to the control given by occurrences of active non single essential or coessential reactions in the SL clusters of the corresponding mutants. \textbf{a Inset}. The same as in \textbf{a} in rich medium.
 \textbf{ d--f}: {\bf Entanglement of SL clusters by overlapping reactions.} Each graph is obtained by considering that two SL clusters are connected by a directed link from SL cluster $i$ to SL cluster $j$ if at least $75\%$ of the reactions in $i$ are also in $j$. Nodes represent SL clusters. The size of a node is proportional to the size of the corresponding SL cluster in number of reactions and the color indicates its maximum k-core out, where a k-core out k-c is defined as the maximal subgraph of SL clusters such that all the SL clusters in it have at least k-c outgoing connections inside the subgraph. Notice that in this representation the internal connections of the highest k-cores may lay beneath the nodes, see Supplementary Fig.~(S4). Each SL cluster is identified by a number, see Supplementary Excel file of each organism for details.}
 \label{fig:3}
\end{figure*}

SL clusters are not disjoint but strongly coalescent in their component reactions and associated genes. The first observation supporting this is given by the repeated usage of the same backup system for different switches in an organism. This redundancy is intensive in {\it E. coli} and {\it S. sonnei}, in which $\approx 20\%$ of the switches share what we call the ``fatty acid biosynthesis backup system'' formed by eleven reactions~\footnote{The ``fatty acid biosynthesis backup system'' includes the four 3-oxoacyl-[acyl-carrier-protein] synthase, the three 3-oxoacyl-[acyl-carrier-protein] reductase reactions, and the four 3-hydroxyacyl-[acyl-carrier-protein] dehydratase reactions in the bacteria.}. This specific redundancy explains the prominent function of Cell Envelope Biosysthesis as a backup for Membrane Lipid Metabolism, with a concentration of PSL pairs at their interface~\cite{Guell:2014b}.

The redundancy not only affects backup reactions. We ranked all reactions participating in SL clusters by their occurrence in different clusters. Results are shown in Fig.~(\ref{fig:3})a-c. We compared the rankings of switches, backups and noncoessential reactions with the control given by the sets of differentially activated reactions resulting when knocking down reactions which are active in WT, but which are not essential or coessential. As expected, the profiles of participation in SL clusters of switches and backups are similar, and differ from the curve of the control test, which decreases much more quickly in {\it E. coli} and {\it S. sonnei} indicating a stronger overlap of SL clusters as compared with rearrangements caused by random deletion of reactions. In line with what observed so far, the picture changes in the case of {\it S. enterica}, with no clear distinction between switches, backup and control reactions. 

As a common trend in the three bacteria, we observe a heterogeneity of participation values such that a small number of reactions participates in a very large fraction of clusters. This suggests that metabolic reorganizations caused by essential plasticity happen mainly by leveraging on some key reactions. For instance, the reductase reactions producing Isopentenyl diphosphate and Dimethylallyl diphosphate, used by organisms in the biosynthesis of terpenes and terpenoids, form a SL cluster in {\it E. coli}~\footnote{The former acts as a switch, producing Isopentenyl diphosphate which is transformed into the more reactive form Dimethylallyl diphosphate by the isopentenyl pyrophosphate isomerization reaction. The reductase reaction producing Dimethylallyl diphosphate acts as a backup.} which coappear in $70\%$ of all its SL clusters. Another intriguing result is related with oxygen consumption in mutants of {\it E. coli} and {\it S. sonnei}. Both organisms display a strong overlap of SL clusters of the mutants which display an altered consumption of oxygen in comparison with WT. Strikingly, all of them share a module of three reactions for the exchange, transport and oxidation of iron, which never participate in SL clusters of mutants which do not show alteration of oxygen consumption. This is in agreement with observations that report oxygen as a signal for regulating iron acquisition in Shigella~\cite{Payne:2007}.

The strong entanglement of SL clusters can also be explored by constructing graphs in which they are represented as nodes and a directed link from cluster $i$ to cluster $j$ indicates that at least $75\%$ of the reactions in $i$ are also in $j$, Fig.~(\ref{fig:3})d-f~\footnote{We have performed other complementary tests, see the network representation of pathways entangled through PSL pairs Supplementary Fig.~(S5) and reaction entanglement matrices showing co-occurrences of reactions in SL clusters Supplementary Fig.~(S6-S8).}. These networks are very densely connected. Their structure can be explored quantitatively using the k-core decomposition~\cite{PhysRevLett.96.040601}, which identifies groups of SL clusters having k-c or more connections among them~\footnote{In complex networks, a k-core of level k-c is defined as the maximal subgraph in which all the nodes have at least k-c connections and can be obtained by repeatedly deleting all nodes with less than k-c connections.}. Each SL cluster is annotated with the maximum value k-c of the k-cores it belongs to, computed on the basis of outgoing connections. Notice that SL clusters in the highest k-cores are very densely connected among them. In {\it E. coli} and  {\it S. sonnei}, the k-core partition denotes a hierarchical core-pheriphery structure. SL clusters are clearly separated into two groups with low and high k-c value, the latter typically formed by the largest SL clusters. The only exception is the size-two SL cluster in {\it E. coli} Isopentenyl- Dimethylallyl diphosphate mentioned above, which acts as an important interface between the core and many peripheral clusters, see Fig.~(\ref{fig:3})d and Supplementary Fig.~(S4). The core contains approximately $1/5$ of the SL clusters and forms an almost fully connected set of reactions participating in many other SL clusters. In contrast, the k-core layout of {\it S. enterica} is almost flat with no relevant core-periphery structure. 

\begin{figure}[t]
 \centering
 \includegraphics[width=8.5cm]{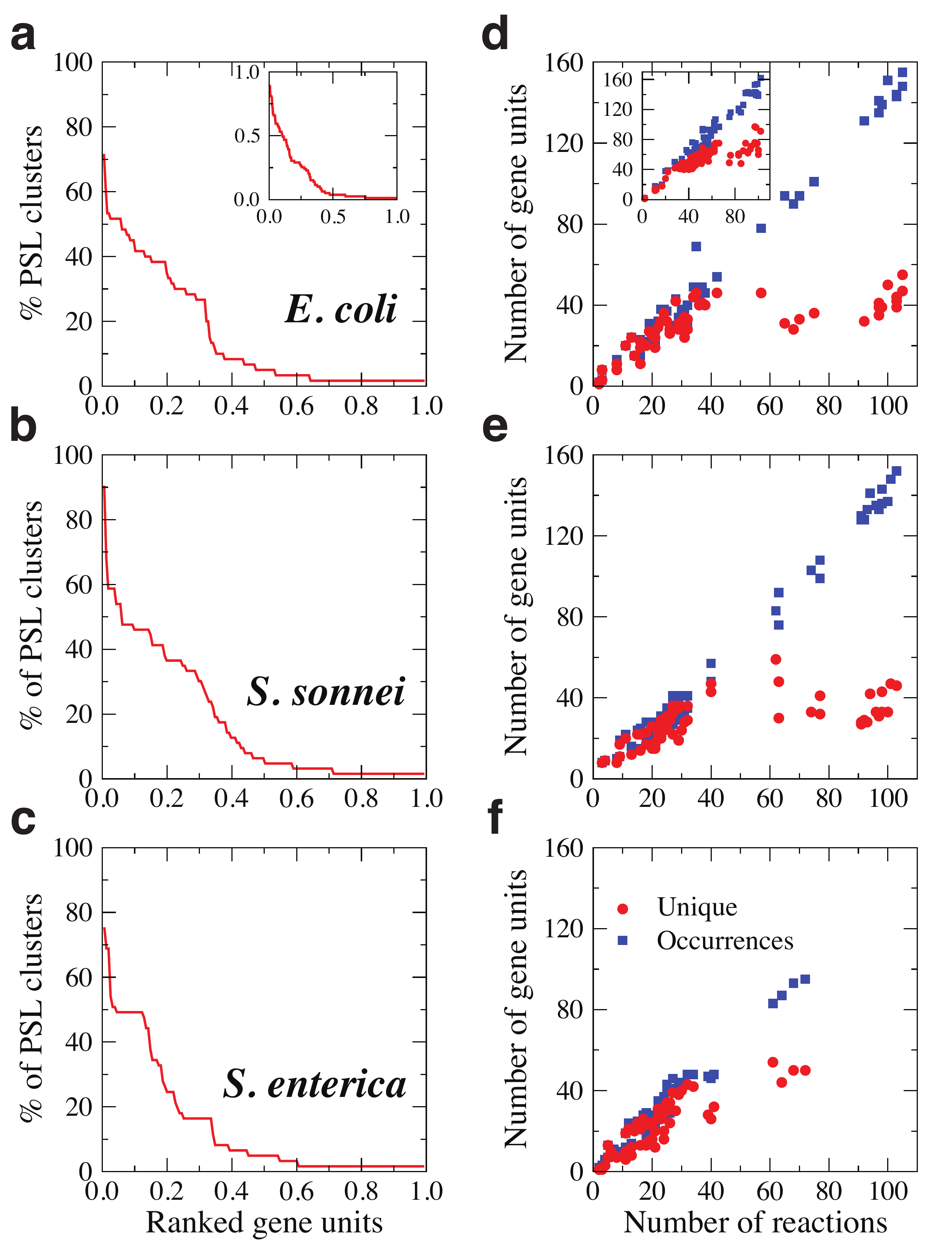}\\
 \caption{\textbf{Redundancy of genes in SL clusters. a-c.} Metabolic gene units (gene and gene complexes) ranked according to the number of SL clusters in which they participate. {\bf d-f} Number of gene units {\it vs} number of reactions in SL clusters. For each organism considered, number of gene units associated to reactions in each SL cluster versus the number of corresponding reactions, including gene repetitions (labeled Occurrences) and excluding them (labeled Unique). }
 \label{fig:4}
\end{figure}

\subsection{Entanglement of SL clusters by overlapping genes}
At the level of genes, the entanglement of SL clusters is even stronger. 
Here, we consider genetic units, which can be single genes or gene complexes (sets of functionally related genes that regulate together a metabolic reaction in a SL cluster via an AND logical relation). To clarify the question whether there is a common set of regulatory genes for SL clusters we ranked metabolic genes according to the number of SL clusters in which they participate, results in Fig.~(\ref{fig:4})a-c. Hub genes, like the one regulating the function of the reductase reactions Isopentenyl and Dimethylallyl diphosphate in {\it E. coli} ~\footnote{This gene appears to be involved in cell lysis and in the stress response of bacteria in reaction to amino-acid starvation, fatty acid limitation, iron limitation, heat shock and other stress conditions. Its action causes the cell to divert resources away from growth and division toward amino acid synthesis in order to promote survival until nutrient conditions improve.}, participate in more than $70\%$ of SL clusters and the top $10\%$ enter in about $50\%$ of the sets. However, more than $50\%$ of the genes is specific to up to three SL clusters. Interestingly, the gene participation curve decays faster for {\it S. enterica} than for the other two bacteria, highlighting again a more limited organization.

In Fig.~(\ref{fig:4})d-f, we plot both the number of unique genes entering in a cluster (unique) and their total number by counting repetitions (occurrences). The number of occurrences grows linearly with the number of reactions in each set, with an approximate slope of $1.4$ in the three bacteria, which indicates that complexes are frequently associated to the regulation of SL clusters. Interestingly, the number of unique genes follows the same linear growth up to a  `critical' value from which it saturates to a constant around $40$ for SL clusters with more than $\sim 60$ reactions (around half the maximum size of SL clusters). The saturation effect implies that large SL clusters are regulated by a reduced number of different genes and that the basin of regulation of genes and the role of complexes grows with the number of reactions in the set. These features are common to the three organisms, although SL clusters in {\it S. enterica} are smaller than in the other two bacteria, so that the saturation effect is less evident and the redundancy of genes is more limited~\footnote{Pairs of genetic units show also a clear tendency to co-occurrence in SL clusters, see the gene entangle matrices for the three bacteria in Supp. Fig.~(S8-10).}. 

\section{{\it E. coli} in rich medium}
Results were also obtained for {\it E. coli} in rich medium (see Methods). The total number of different reactions in SL clusters is a factor $1.8$ larger than in minimal medium and the average number of backups per switch increases approximately in the same proportion. The number of switches is slightly increased, see Table I. Similar to glucose minimal medium, the size distribution of SL clusters has a longer tail as compared with the control, inset Fig.~(\ref{fig:1})b, and SL clusters of differentially activated reactions are formed basically by a connected component, inset Fig.~(\ref{fig:1})e. Of the $287$ reactions in SL clusters in glucose minimal medium, $85\% (244)$ also belong to SL clusters in rich medium. Of them, only $8.6\% (21)$ change role. In particular, $15$ which were coessential are rescued and become nonessential in rich medium. More than $70\%$ of switches and backups in glucose minimal medium are conserved, $43$ and $42$ respectively. Of them, $40$ switches preserve exactly the same backup system, $2$ switches acquire an extra backup reaction, and $1$, corresponding to the switch reaction isopentenyl pyrophosphate isomerization~\footnote{The isomers Isopentenyl diphosphate, less reactive, and Dimethylallyl diphosphate, more reactive, can only be produced by the corresponding reductase reaction and by the isopentenyl pyrophosphate isomerization reaction. In glucose minimal medium, the more reactive form is obtained via the less reactive isomerase and not directly by the action of the corresponding reductase, so that the reductase producing Isopentenyl diphosphate and the isomerization reaction act as switches with the reductase reaction producing directly Dimethylallyl diphosphate as a backup. The reverse is observed in rich medium, where the less reactive form is obtained from the more reactive one so that the reductase producing Dimethylallyl diphosphate and the isomerization reaction act as switches with the reductase reaction producing directly Isopentenyl diphosphate as a backup.}, changes backup. Interestingly, in two more SL cluster the switch and the backup swap roles so that alternative mechanisms are used to produce specific metabolites~\footnote{One of them is related to the production of deoxyuridylic acid, an intermediate in the metabolism of deoxyribonucleotides. The hidrolase reaction, without the direct intervention of ATP, is active in glucose minimal medium the active reaction in rich medium is the thymidine kinase catalysed reaction, which involves the direct consumption of ATP}. 

We see that the cost and energetic requirements of rearrangements from WT to mutant do vary between minimal and rich medium conditions (Fig. \ref{fig:2}). The flux ratio per module $F^{\rm mutant}/F^{\rm WT}$ tends in fact to generally decrease, at odds with the minimal medium case. This is however reasonable, since in the rich medium there are  several metabolic routes that connect nutrients to biomass. Introducing mutations may disrupt many of these routes (literally switching off the metabolism of some of the redundant nutrients) without impairing survival, decreasing the total flux running through the modules as an effect. In the minimal medium case, instead, it is impossible to disrupt these routes without incurring in lethality. Besides this discrepancy, switches and backup participations in different SL clusters decrease faster than non coessential and control reactions, suggesting that SL cluster entanglement decreases dramatically in the presence of multiple nutrients, as also suggested by the reduced average number of backups per switch.

\section{Discussion}\label{sec:discussion}
Metabolic networks can change their state significantly without causing the loss of an organism's ability to survive in a given environment, and this property allows it to explore a wide range of novel metabolic abilities~\cite{Rodrigues:2009}. However, flux rerouting has been claimed as negligible in providing robustness for a large number of mutant strains~\cite{Papp:2004,Blank:2005,Wunderlich:2006}. Nevertheless, metabolic flux reorganization becomes essential in some critical situations, in particular when reactions in synthetic lethal pairs fail. The big majority of SL interactions in the studied bacteria involve the activation of silent backup coessential partners to ensure the viability of the mutants. This essential plasticity is mediated by the activation and inactivation of reactions in SL clusters acting as backup systems containing coessential but also nonessential or coessential reactions, in contrast to the model of SL interactions restricted to coessential pairs. 

The robustness that confers essential plasticity comes at the expenses of acquired vulnerabilities and of increased structural and functional costs. These costs are manifested in an increased number of active reactions and total flux running through SL clusters, lower efficiency in energetic production, a slightly reduced biomass yield, and an increased centrality of reactions in the backup systems. We hypothesize that, despite the increased functional and structural cost of viability in SL mutants, the expected burden for sustaining alternative backup systems for all switch reactions in the organism is indeed buffered by the overlap of SL clusters. This is supported by our observation that SL cluster entanglement decreases markedly in the presence of multiple nutrients. The regulation of small sets of enzyme-coding genes controlling a reduced number of differentially activated reactions which participate in many different SL clusters is sufficient to provide the varied combinations necessary to protect the bacterium against a diversity of switch mutations. We believe that the higher centrality of backups, which act as a sort of local hubs in SL clusters, is related to the requirement of an efficient regulation of metabolism --e.g., by transcriptional regulation~\cite{DeRisi:1997}-- controlling the alternative metabolic flux routing within the network in order to sustain viability in the event of a harmful mutation, like the knockout of a switch. This comes at the expenses of an increased vulnerability of the organisms in the new mutated state, since the increased centrality of backup reactions as compared to switches makes the metabolic structure of the mutant more fragile and vulnerable to potential failures~\cite{Jeong:2001}, and suggests a tradeoff between robustness and the efficiency of backup regulation.

The existence of SL clusters and their strong entanglement in minimal medium is a common feature in the three bacteria analysed here. The entanglement of SL clusters is observed at two different levels. First, silent PSL coessential reactions tend to form leagued backup systems which are shared by several switches. Second, clusters are also entangled by other noncoessential reactions. However, some specific features seem more sensitive to evolutionary pressure, as revealed by a comparative analysis of the results for the three species. In fact, {\it E. coli} and {\it S. sonnei} are extremely congruent, in accordance with the very short phylogenetic distance separating them in the evolutionary tree and with previous results claiming members of the genus Shigella as ``{\it Eschericia coli} in disguise''~\cite{Lan:2002}. Differences are however interesting, like the patterns of oxygen consumption observed in a high proportion of SL mutants versus WT. The increase in {\it E. coli} implies a reduced efficiency in energy production by oxidative processes. The decrease in {\it S. sonnei} denotes a change of strategy in the production of energy from aerobic to anaerobic mechanisms. On the other hand, {\it S. enterica} shows a differentiated profile, less complex and structured, in line with its role as evolutionary ancestor. The results obtained in minimal medium suggest that {\it E. coli} and {\it S. sonnei} are more optimized towards growth and biomass yield and undergo a much more complex metabolic reorganization to face mutations and still achieve performances comparable to the WT. 

Essential plasticity, a more sophisticated mechanism as compared to redundant metabolic flows in other SL interactions, seems to be promoted by evolutionary pressure but, at the same time, tends to increase the degeneracy and centrality of backup systems as a regulatory mechanism ensuring the entanglement of SL clusters forming a hierarchical core-periphery structure. This entanglement economizes the huge potential metabolic burden due to the maintenance  alternative metabolic routes for all PSL interactions in an organism. On the other hand, the flux reorganization of {\it E. coli} in rich medium, in which cluster entanglement decreases markedly, lead us to think that the strength of entanglement can be responsive to environmental stresses, like starvation. 

In summary, we propose a change of paradigm in the approach to understand the phenomenon of synthetic lethality. The complexity of molecular interactions at the cell level urge us to go from the mere screening of SL reaction or gene pairs, or even of triplets or higher order motifs, to the study of SL clusters and their entanglement. Approaching directly SL pairs of reactions or genes without their multifunctional integration in clusters is like drawing paths between pairs of geographical places without the scaffold of a map telling how the different paths relate to each other. The complete portray at the systems level is far more complex than a collection of separate PSL pairs. Beyond theoretical implications for the understanding of plasticity in metabolic networks, our results could help to identify drug action and to design improved strategies that reduce undesired resistance in synthetic lethal interactions to chemicals in pathogens. We believe that SL clusters will be also found in human cells, with important implications for biomedicine and biotechnology. Our work reveals that not all SL pairs have the same quality as potential therapeutic targets in complex diseases such as cancer or to fight infections of pathogens. We expect that more redundant coessential reactions with a higher participation in different SL clusters can become efficient and reliable supertargets. 

\section{Methods}

\subsection{Statistics of the genome-scale metabolic networks}
For each of the three bacteria, we report the metabolic reconstruction (Organism, Model, Reference), the number of reactions included in the genome-scale reconstruction ($N_R$ all), the number of reactions possibly active according to FVA ($N_R$ FVA), the number of active reactions in the FBA solution in glucose minimal medium ($N_R$ active), the number of metabolites in the reconstruction ($N_M$ actual) and those resulting when considering compartments ($N_M$ synth), the number of single essential reactions, the number of Plasticity SL pairs, and of Redundancy SL pairs, the number of SL clusters and switches ($R_{switches}$), and the number of backups ($R_{backups}$), noncoessential reactions ($R_{noncoess}$), total different reactions (Total $R$), and Gene units associated to the SL clusters. We also report the ratio of the total flux running through reactions in SL clusters of mutants and WT ($[F_{mutant}/F_{WT}]_{cluster}$), the ratio of active reactions in SL clusters of mutants and in the WT ($A_{NG,mutant}/A_{WT}$), and the ratio of ATP production in PSL mutants and in the WT ($E_{NG,mutant}/E_{NG,WT}$). In relation to centrality measures, we report the fraction of times that backups have higher centrality than switches inside the internal connected component of SL clusters $C_{BgtS}$, the fraction of times that backups have higher centrality than non coessential reactions inside the internal connected component of SL clusters $C_{BgtNC}$, the average centrality of backup reactions in the internal connected component of SL clusters $\left<C_B \right>$, the average centrality of switch reactions in the internal connected component of SL clusters $\left<C_S \right>$, the average centrality of noncoessential reactions in the internal connected component of SL clusters $\left<C_NCE \right>$. For {\it E. coli}, values in parentheses correspond to rich medium, see next subsection.
\begin{table}
\caption {Metabolic reconstructions considered in our study and statistics in glucose minimal medium.}
\label{tab:recons}
\begin{tabular}{|l|c|c|c|c|}
\hline
 \multicolumn{2}{|c|}{\multirow{2}{*}{Organism}}  &{\it E. coli}  & {\it S. sonnei}&{\it S. enterica} \\
 \multicolumn{1}{|c}{}&&MG1655 & Ss046 & LT2 \\\hline
 \multicolumn{2}{|c|}{Model} &{\it i}JO1366&  {\it i}SSON$\_$1240 &STM$\_$v1$\_$0\\  \hline 
  \multicolumn{2}{|c|}{Reference} &\cite{Orth:2011a}&  \cite{Monk:2013a} &\cite{Thiele:2011} \\  \hline \hline
 \multirow{4}{*}{$N_R$}&all&2583&2694& 2546\\\cline{2-5}
&unlocked&2284&2334& 2227\\\cline{2-5}
&FVA& 1520 & 1757 & 1678 \\\cline{2-5}
&active&418&450& 498\\\hline\hline
 \multirow{2}{*}{$N_M$}&actual& 1136&1216&1119 \\\cline{2-5}
&synth& 1805&1938& 1802\\\hline\hline
 \multicolumn{2}{|c|}{Single Essential} &289 & 282 &346\\\hline\hline
 \multicolumn{2}{|c|}{Plasticity SL}& 231&233& 109\\\hline\hline
 \multicolumn{2}{|c|}{Redundancy SL} &25 & 30 &31\\\hline\hline
  \multirow{5}{*}{PSL}&$R_{switches}$&60 (79)&63&61\\\cline{2-5}
&$R_{backups}$&59 (136)&60&63\\\cline{2-5}
&$R_{noncoess}$&168 (293)&152&116\\\cline{2-5}
clusters&Total $R$&287 (508)&275&240 \\\cline{2-5}
&Gene units&196&161&155 \\\hline\hline
  \multicolumn{2}{|c|}{$[F^{mutant}/F^{WT}]_{cluster}$} &542.69 & 6.56 &90446.79\\\hline
    \multicolumn{2}{|c|}{$A^{mutant}/A^{WT}$} &1.56 & 176.14 &0.64\\\hline
    \multicolumn{2}{|c|}{$E_{NG}^{mutant}/E_{NG}^{WT}$} &15.56 & 1.25 &1.25\\\hline\hline
          \multicolumn{2}{|c|}{$C_{BgtS}$} &0.64 & 0.64 &0.37\\\hline
 \multicolumn{2}{|c|}{$C_{BgtNC}$} &0.56 & 0.55 &0.49\\\hline
       \multicolumn{2}{|c|}{$\left<C_B \right>$} &0.50 & 0.48 &0.58\\\hline
 \multicolumn{2}{|c|}{$\left<C_S \right>$} &0.43 & 0.44 &0.55\\\hline
    \multicolumn{2}{|c|}{$\left<C_{NCE} \right>$} &0.41 & 0.40 &0.48\\
\hline
\end{tabular}
\end{table}

\subsection{Flux Balance Analysis and environmental conditions}
Flux Balance Analysis (FBA)~\cite{Orth:2010a} is a technique which allows to compute metabolic fluxes without the need of kinetic parameters, just by using constrained-optimization. The vector of the time variation of the concentrations of metabolites $\dot{c}$ is related with the stoichiometric matrix $S$ of the whole network (it contains the stoichiometric coefficients of each metabolite in each reaction of the network) and the vector of fluxes $\nu$, $\dot{c}=S\cdot \nu$. Steady-state is assumed, thus $S \cdot \nu=0$. In general, metabolic networks contain more reactions than metabolites, and hence the system of equations for the fluxes is underdetermined. Hence, a biological objective function must be defined in order to select a biologically meaningful solution. In this work, we use FBA to find the solution that optimizes the growth of the organism, which is equivalent to maximize biomass formation. Reversibility of reactions is also added in order to constrain the solutions. Since we have a linear system of equations with linear constraints, Linear Programming is used in order to compute a flux solution in a small amount of time (of the order of 1 s), which implies a computationally cheap method. According to the specifications in each metabolic reconstruction, growth in glucose minimal medium was simulated by fixing the lower bound of the glucose exchange reaction to $-10  mmol/(gDW\cdot h)$ for {\it E. coli} and {\it S. sonnei}, and to $-5  mmol/(gDW\cdot h)$ for {\it S. enterica}.

As the rich medium for {\it E. coli} we used a Luria-Bertani Broth~\cite{Sezonov:2007a}, which contains as additional compounds purines and pirimidines apart from amino acids. We also added vitamins, namely biotin, pyridoxine, and thiamin, and also the nucleotide nicotinamide monocleotide~\cite{Wunderlich:2006}. Other compounds, like PABA or chorismate, cannot be uptaken by the {\it E. coli}  model that we are using. The exchange constraints bounds of these compounds are set to $-10 mmol/(gDW\cdot h)$ $(\nu_{exchange}^{compound}\ge -10)$. A detailed list of the added compounds is given in our Supplementary Data Table S4.

\subsection{Detection of SL pairs}
From the set of reactions in the genome-scale reconstructions, we excluded essential reactions detected computationally and also spontaneous reactions (e.g transport reactions). We focused exclusively on reactions catalyzed by enzymes with an associated gene. In this way, we identified a set of candidate reactions in each organism that can be removed individually, but whose pair deletion may be lethal for the organism. We checked every possible pair by applying FBA to the double mutant. As in~\cite{Guell:2014b}, we classified the detected SL pairs into plastic and redundant, depending on whether only one or both reactions are active in the FBA solution in the given medium.

\subsection{Computation of the centrality of reactions in SL clusters}
We modelled metabolism as a bipartite directed network~\cite{Guell:2012}, where directed links connect metabolites with reactions in which they participate as reactants or products. The degree centrality of a reaction is simply given by its degree $k$, measuring the number of other reactions connected to it by shared metabolites. We define the normalized degree centrality of a reaction in the internal connected component of a SL cluster as $k_r/(R-1)$, where $k_r$ stands for the number of reactions connected to reaction $r$ inside the connected component, and $R$ is its total number of reactions.

\section*{Acknowledgments}
F.A.M. acknowledges financial support from the grant PTQ-14-06718 by the Torres Quevedo programme of the MINECO. F.S. acknowledges support from MINECO projects no. FIS2013-41144P and FIS2016-78507-C2-1-P (AEI/FEDER, UE). M.A.S. acknowledges support from the James S. McDonnell Foundation Scholar Award in Complex Systems; MINECO projects no. FIS2013-47282-C2-1-P and no. FIS2016-76830-C2-2-P (AEI/FEDER, UE); and the Generalitat de Catalunya grant no. 2014SGR608. 

\section*{Author contributions}
F.~A.~M., F.~S., and M.\'A.S. contributed to the design and implementation of the research, to the analysis of the results, and to the writing of the manuscript.


\newpage
\onecolumngrid
\section{Supplementary Figures Metabolic plasticity in synthetic lethal mutants: viability at higher cost}
\setcounter{page}{1}
\renewcommand{\thepage}{S\arabic{page}} 
\renewcommand{\thetable}{S\arabic{table}}  
\renewcommand{\thefigure}{S\arabic{figure}} 
\renewcommand{\theequation}{S\arabic{equation}}
\begin{figure*}[h!]
 \centering
 \includegraphics[width=0.6\textwidth]{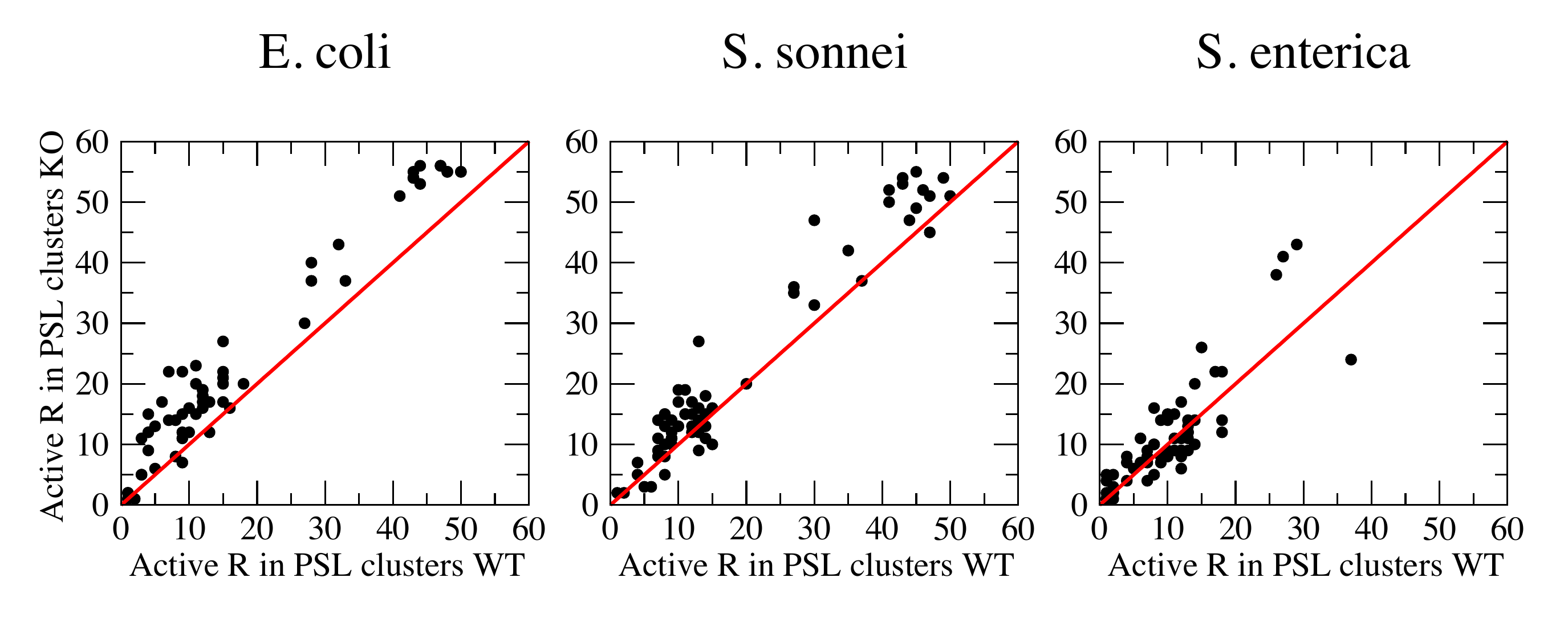}
 \caption{\textbf{Number of active reactions in PSL clusters of mutants versus WT for the three bacteria.} Scatter plots in which each dot represents a PSL cluster and the diagonal line denotes equal number of active reactions.}
 \label{fig:Supp1}
\end{figure*}

\begin{figure*}[h]
 \centering
 \includegraphics[width=0.6\textwidth]{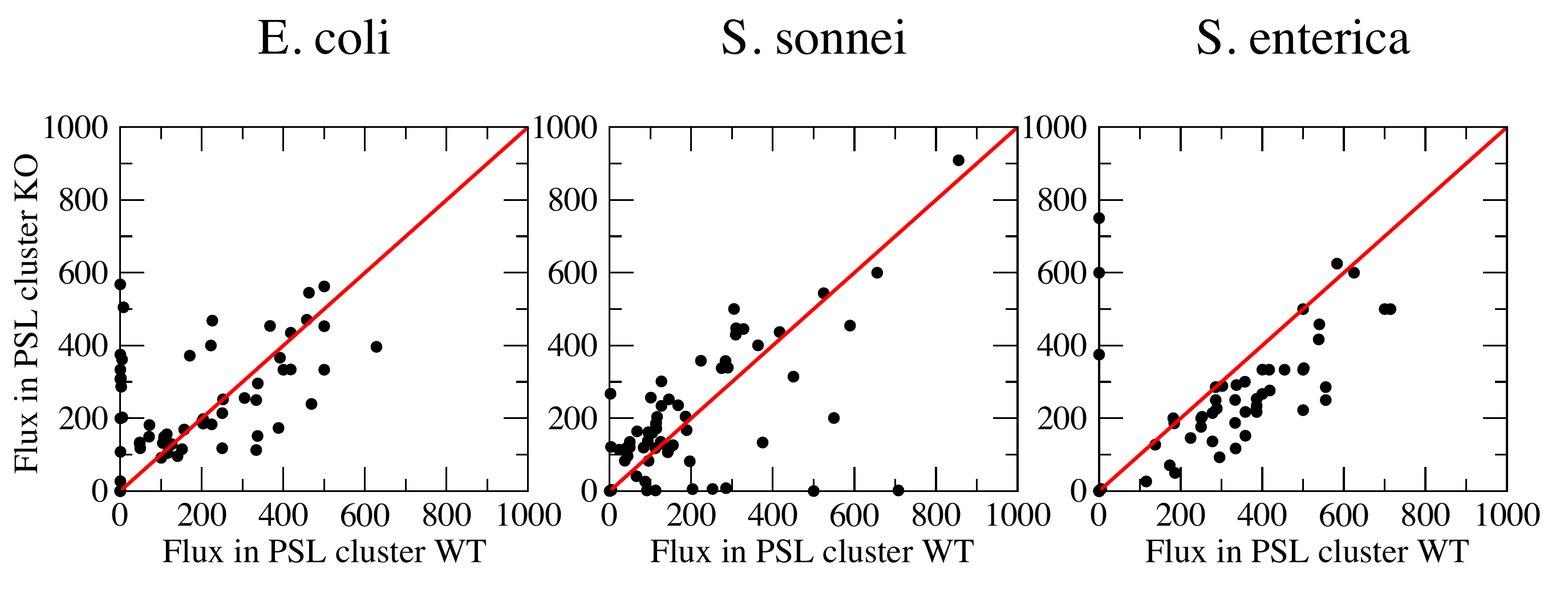}
 \caption{\textbf{Average flux per active reaction in PSL clusters of mutants versus WT for the three bacteria.} Scatter plots in which each dot represents a PSL cluster and the diagonal line denotes equal average fluxes.}
 \label{fig:Supp2}
\end{figure*}

\vspace{0.5cm}
\begin{figure*}[h]
 \centering
 \includegraphics[width=0.8\textwidth]{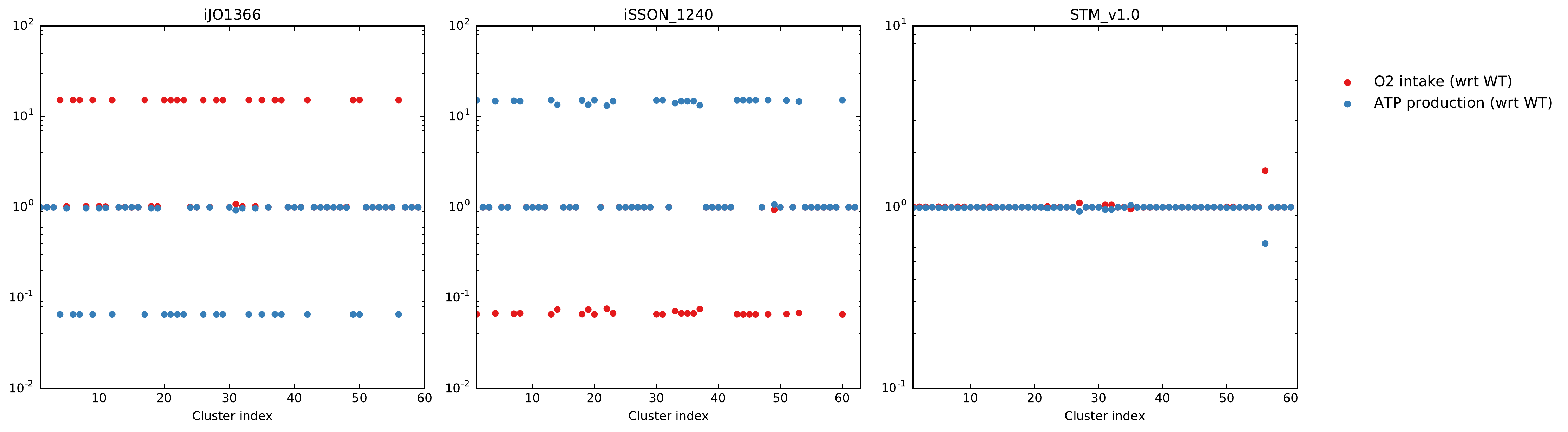}
 \caption{\textbf{ATP production and oxygen consumption of mutants versus WT for the three bacteria.} Each dot represents a PSL cluster.}
 \label{fig:Supp2}
\end{figure*}

\begin{figure*}[t]
 \centering
 \includegraphics[width=1\textwidth]{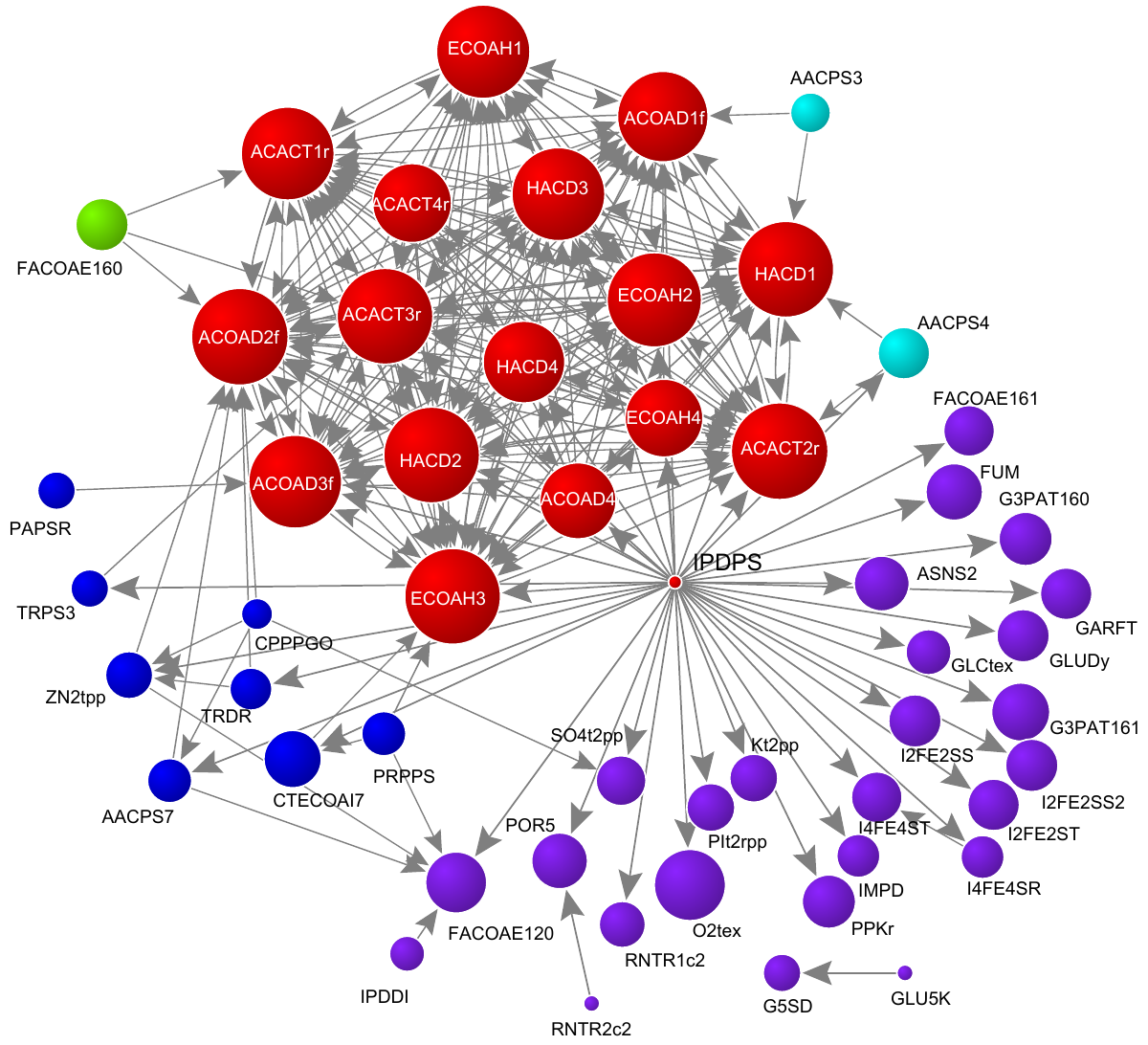}
 \caption{{\bf Entanglement of PSL clusters by overlapping reactions in {\it E. coli}.} The graph is obtained by considering that two PSL clusters are connected by a directed link from PSL cluster $i$ to PSL cluster $j$ if at least $75\%$ of the reactions in $i$ are also in $j$. Nodes represent PSL clusters. The size of a node is proportional to the size of the corresponding PSL cluster in number of reactions and the color indicates its maximum k-core out, where a k-core out $k-c$ is defined as the maximal subgraph of PSL clusters such that all the PSL clusters in it have at least $k-c$ outgoing connections inside the subgraph. }
 \label{fig:Supp3}
\end{figure*}

\begin{figure*}[t]
 \centering
 \includegraphics[width=1\textwidth]{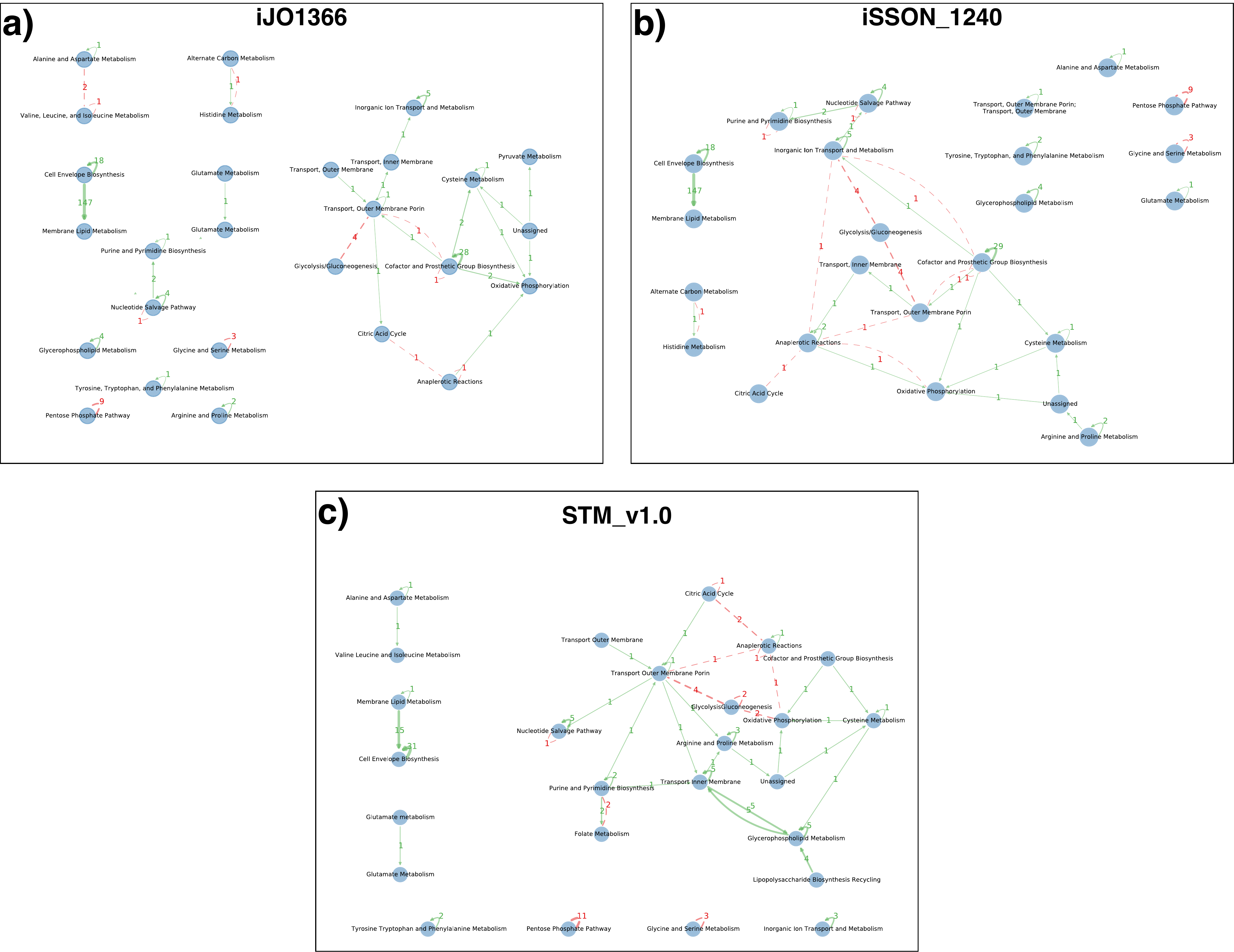}
 \caption{\textbf{Back-up cell envelope-membrane lipid for the three types of bacteria.}  We annotated reactions in PSL pairs in terms of biochemical pathways to provide a network representation where pathways are nodes linked whenever they participate together in a PSL interaction. The network summarizes which pathways are, overall, a metabolic backup of others, for the case of \textbf{(a)} {\it E. coli}, \textbf{(b)} {\it S. sonnei} and \textbf{(c)} {\it S. enterica}. We observe that PSL pairs happen mostly intra-pathway, with the striking exception of the strong entanglement between Cell Envelope Biosysthesis and Membrane Lipid Metabolism in {\it E. coli} and  {\it S. sonnei}. }
 \label{fig:Supp4}
\end{figure*}

\clearpage
\begin{figure*}[t]
 \centering
 \includegraphics[width=1.0\textwidth]{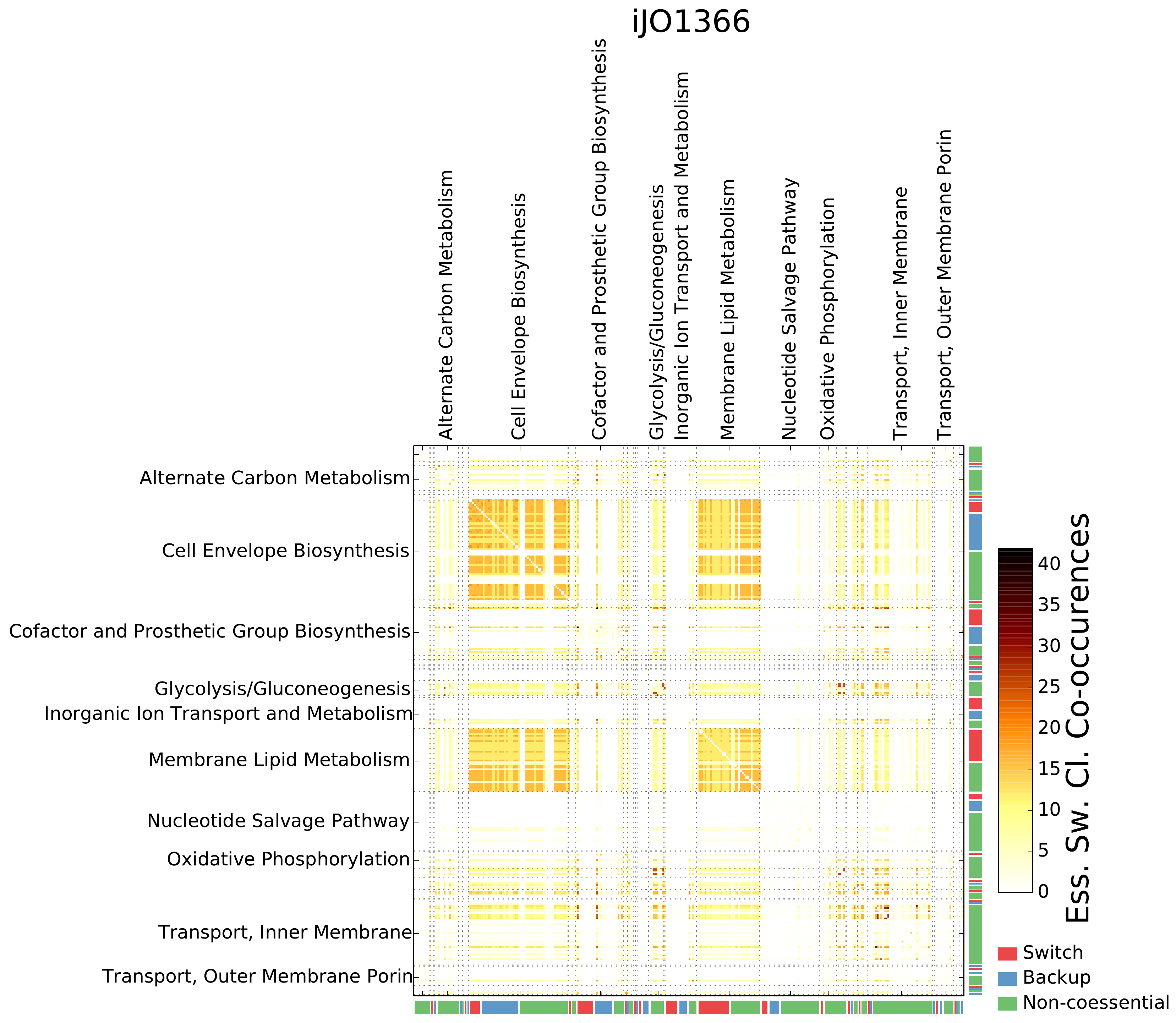}
 \caption{\textbf{Entanglement matrix for pairs of reactions in PSL clusters of {\it E. coli}}. Each matrix shows the number of clusters in which a pair of reactions in PSL clusters coappear. Reactions are ordered according to their metabolic pathway. Pathways are numbered and reported in the list below the matrices. The type of reactions (switch, backup, noncoessential) in the pair is denoted by the color key besides the matrices, e.g. switch reactions are in red. Each entry in the matrix corresponds to the number of PSL clusters in which the corresponding pair of reactions coappear.}
 \label{fig:Supp5}
\end{figure*}

\clearpage

\begin{figure*}[t]
 \centering
 \includegraphics[width=1.0\textwidth]{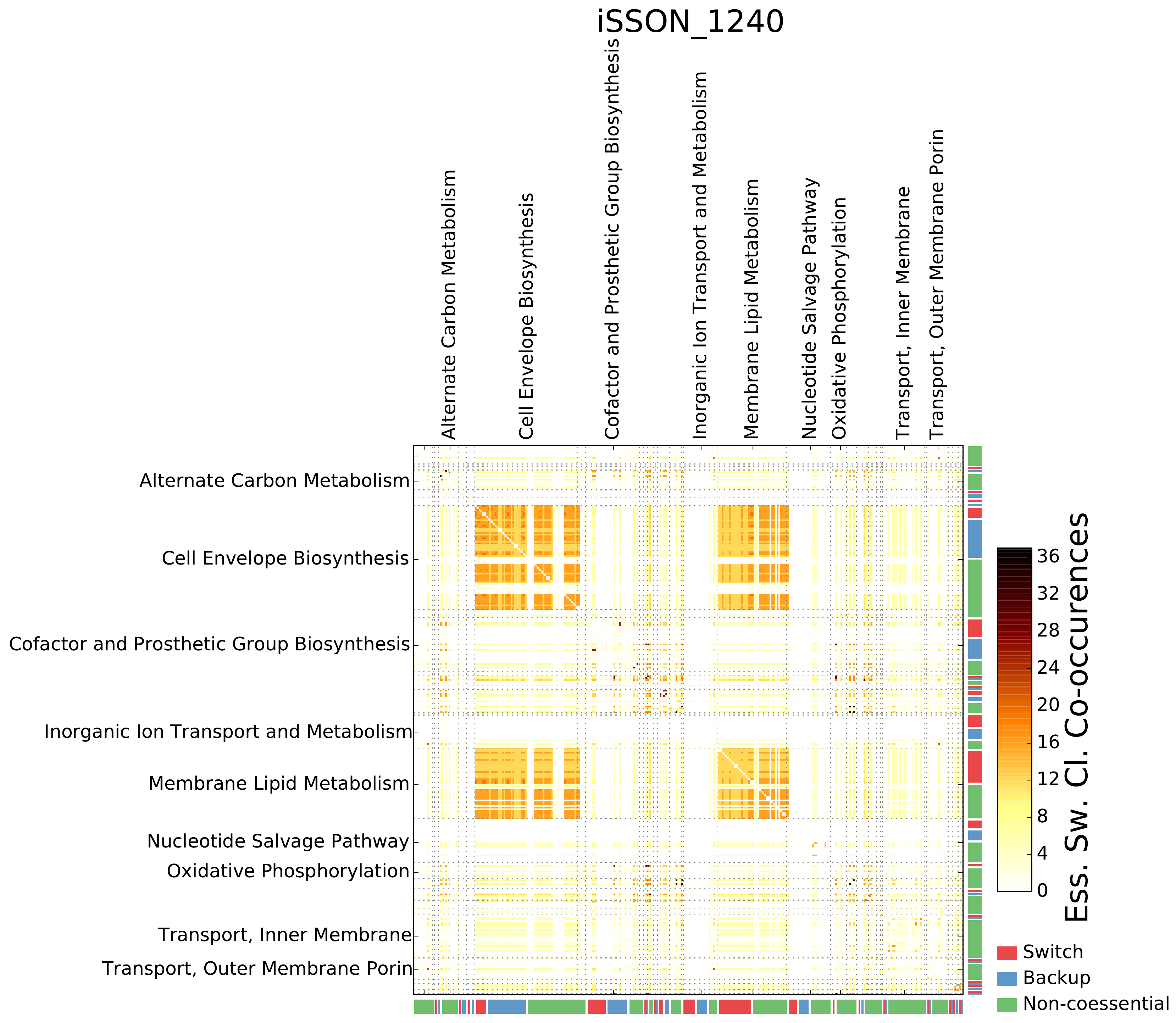}
 \caption{\textbf{Entanglement matrix for pairs of reactions in PSL clusters of {\it S. sonnei}}. Each matrix shows the number of clusters in which a pair of reactions in PSL clusters coappear. Reactions are ordered according to their metabolic pathway. Pathways are numbered and reported in the list below the matrices. The type of reactions (switch, backup, noncoessential) in the pair is denoted by the color key besides the matrices, e.g. switch reactions are in red. Each entry in the matrix corresponds to the number of PSL clusters in which the corresponding pair of reactions coappear.}
 \label{fig:Supp6}
\end{figure*}

\clearpage

\begin{figure*}[t]
 \centering
 \includegraphics[width=1.0\textwidth]{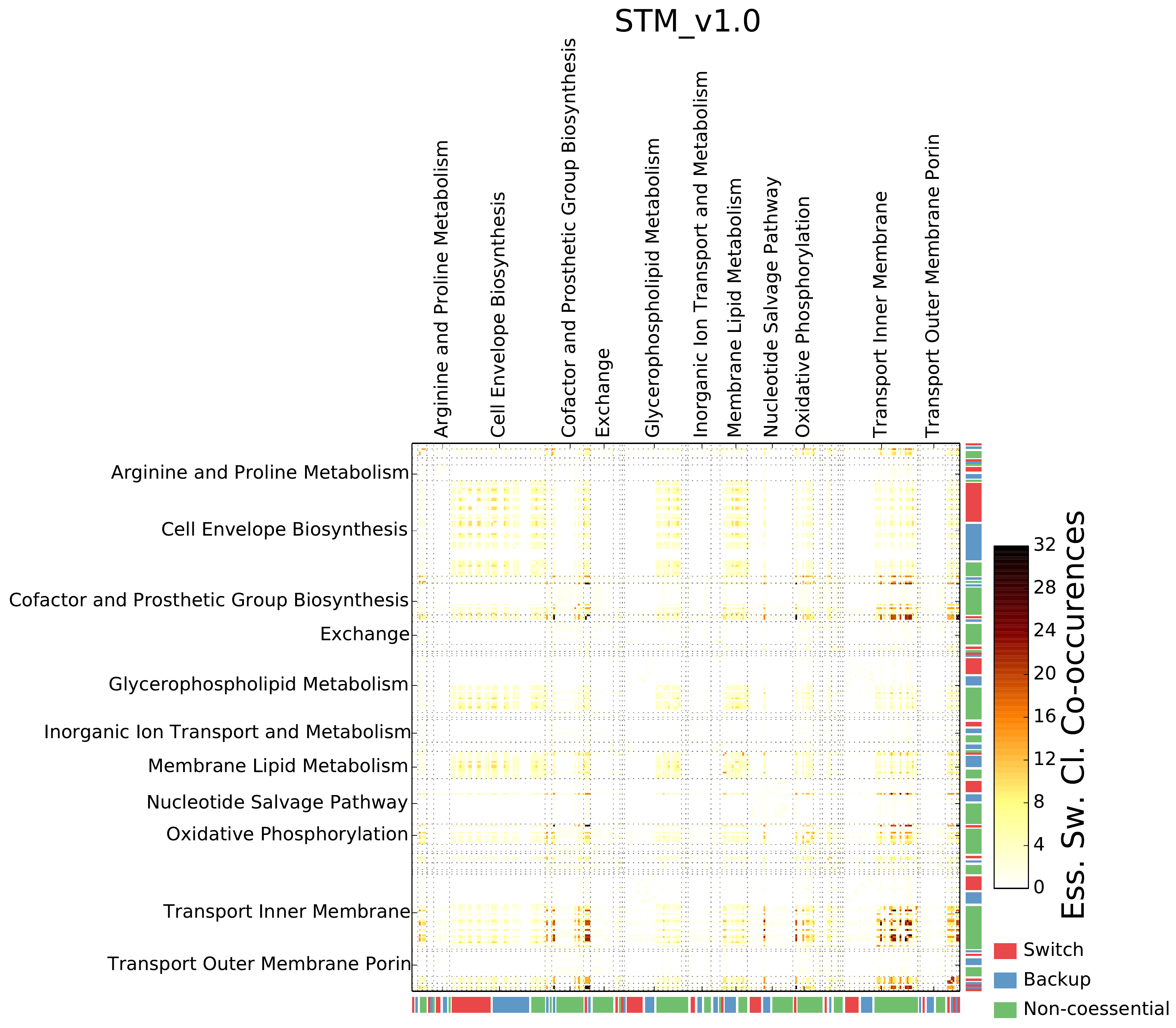}
 \caption{\textbf{Entanglement matrix for pairs of reactions in PSL clusters of {\it S. enterica}}. Each matrix shows the number of clusters in which a pair of reactions in PSL clusters coappear. Reactions are ordered according to their metabolic pathway. Pathways are numbered and reported in the list below the matrices. The type of reactions (switch, backup, noncoessential) in the pair is denoted by the color key besides the matrices, e.g. switch reactions are in red. Each entry in the matrix corresponds to the number of PSL clusters in which the corresponding pair of reactions coappear.}
 \label{fig:Supp7}
\end{figure*}


\clearpage

\begin{figure*}[t]
 \centering
 \includegraphics[width=1.0\textwidth]{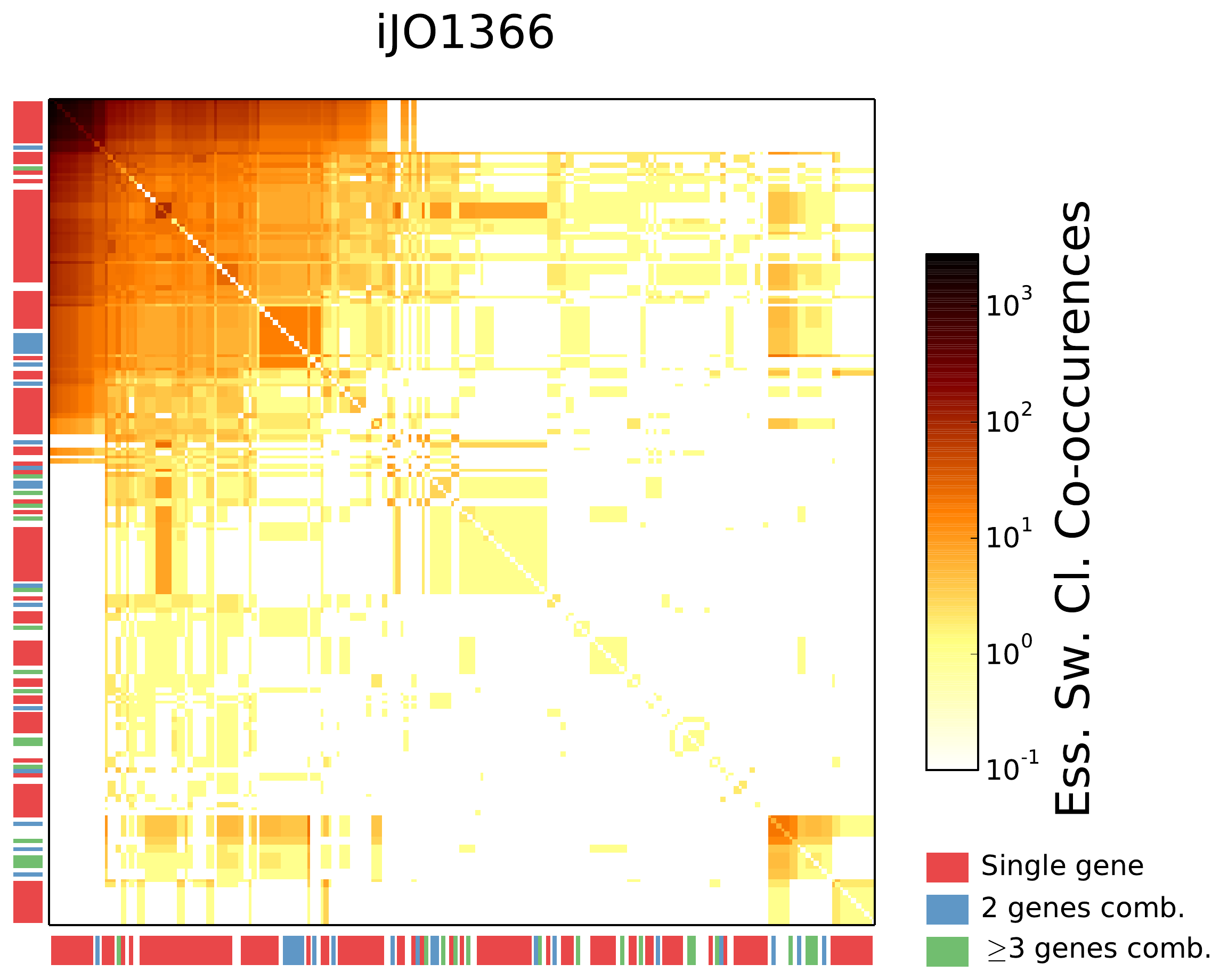}
 \caption{\textbf{Entanglement matrix for pairs of genes or gene complexes in PSL clusters of {\it E. coli}}. Each matrix shows the number of sets in which a pair of genetic units (gene or gene complexes) in PSL clusters coappear. Genetic units are ordered making use of the Infomap algorithm \cite{Rosvall:2008}. Each complex is composed of a number of genes varying from 1 up to 13 and may appear more than once in each set. For this reason, pairs of gene complexes may have a cooccurrence frequency that exceeds the number of sets, as it can be observed mostly in the upper diagonal part of the matrices. The number of genes in the complex is denoted by the color key beside the matrix (e.g. red denotes single genes).}
 \label{fig:Supp8}
\end{figure*}

\clearpage

\begin{figure*}[t]
 \centering
 \includegraphics[width=1.0\textwidth]{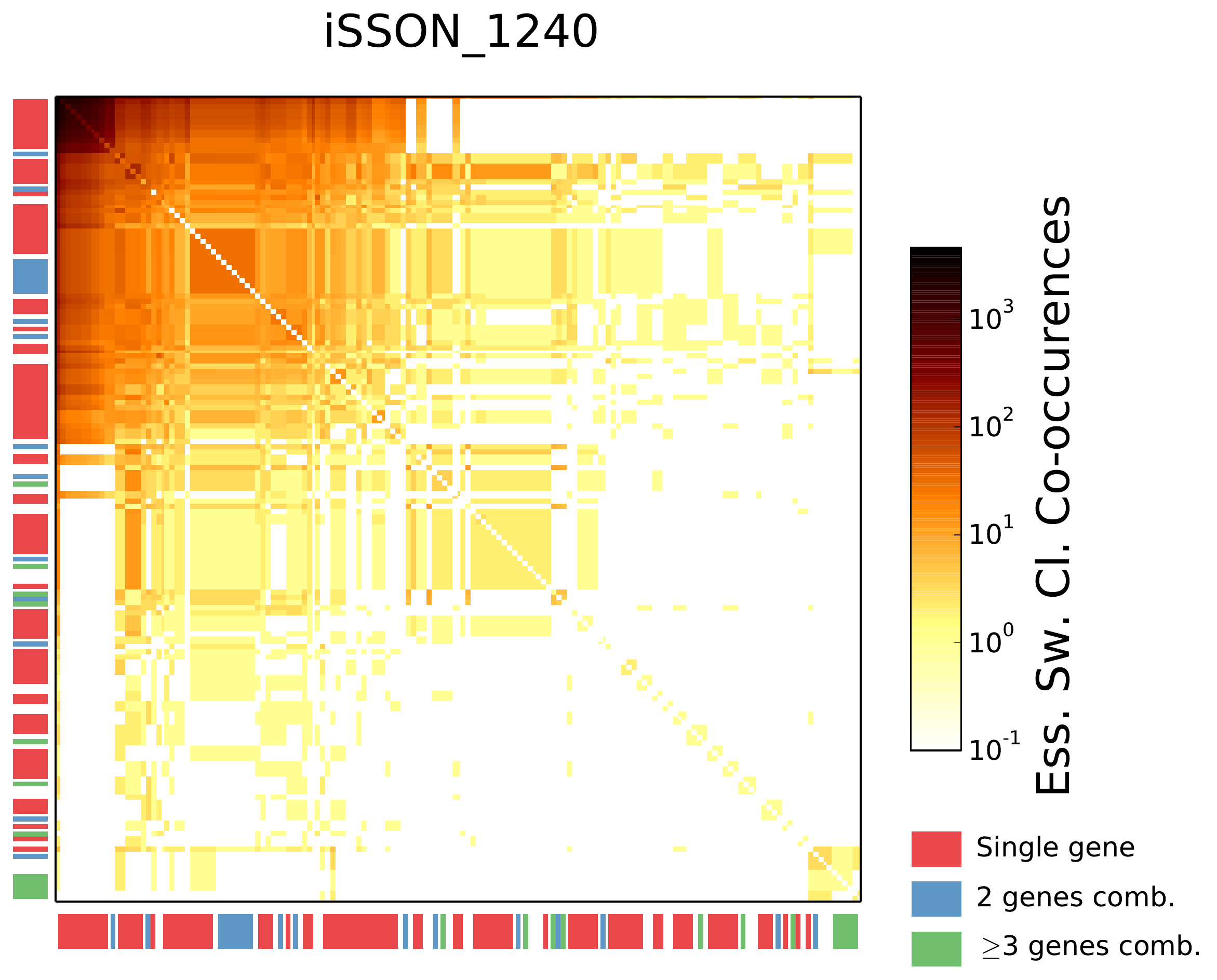}
 \caption{\textbf{Entanglement matrix for pairs of genes or gene complexes in PSL clusters of {\em S. sonnei}}. Each matrix shows the number of sets in which a pair of genetic units (gene or gene complexes) in PSL clusters coappear. Genetic units are ordered making use of the Infomap algorithm \cite{Rosvall:2008}. Each complex is composed of a number of genes varying from 1 up to 13 and may appear more than once in each set. For this reason, pairs of gene complexes may have a cooccurrence frequency that exceeds the number of sets, as it can be observed mostly in the upper diagonal part of the matrices. The number of genes in the complex is denoted by the color key beside the matrix (e.g. red denotes single genes).}
 \label{fig:Supp9}
 
 \clearpage
 
\end{figure*}
\begin{figure*}[t]
 \centering
 \includegraphics[width=1.0\textwidth]{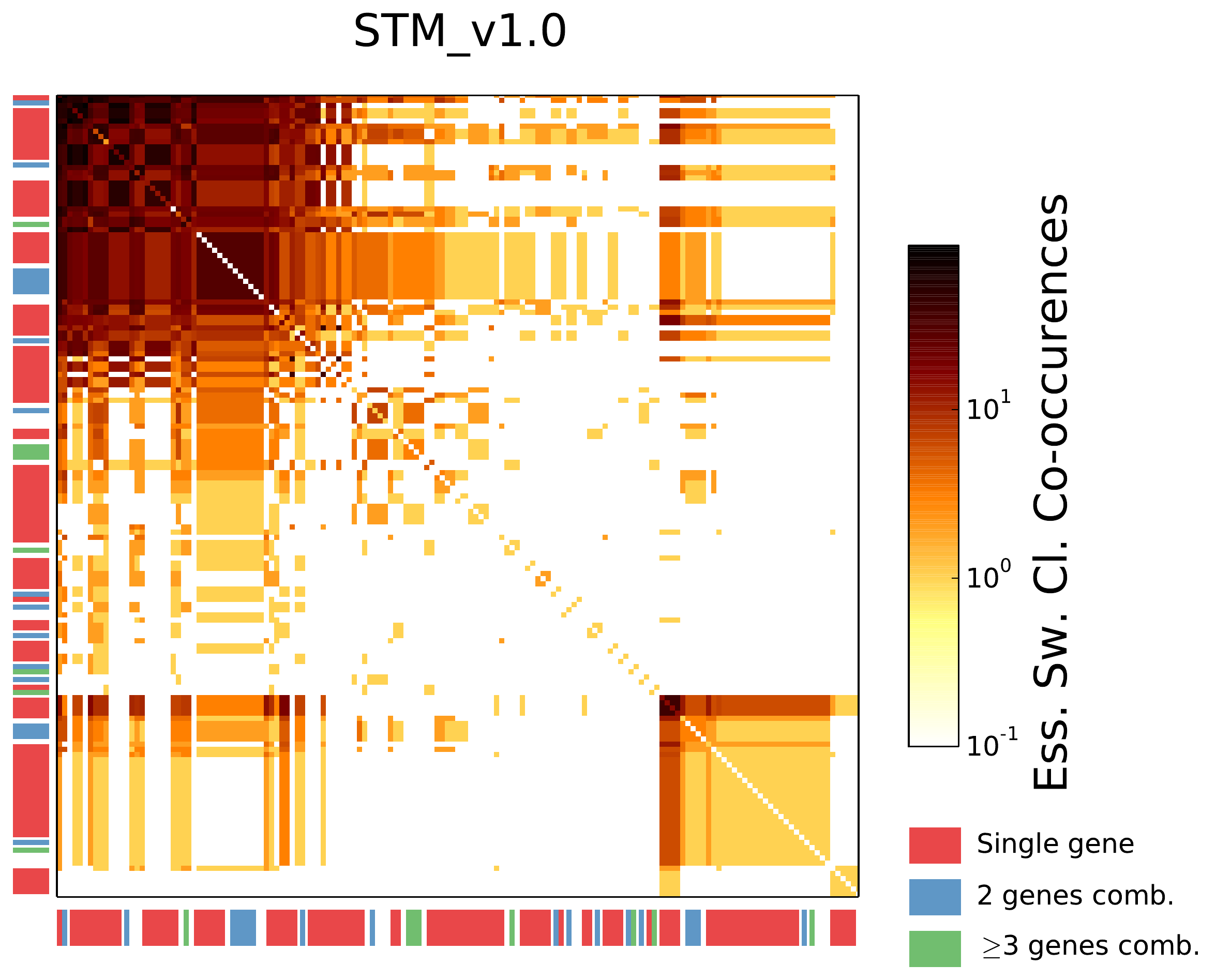}
 \caption{\textbf{Entanglement matrix for pairs of genes or gene complexes in PSL clusters of {\it S. enterica}}. Each matrix shows the number of sets in which a pair of genetic units (gene or gene complexes) in PSL clusters coappear. Genetic units are ordered making use of the Infomap algorithm \cite{Rosvall:2008}. Each complex is composed of a number of genes varying from 1 up to 13 and may appear more than once in each set. For this reason, pairs of gene complexes may have a cooccurrence frequency that exceeds the number of sets, as it can be observed mostly in the upper diagonal part of the matrices. The number of genes in the complex is denoted by the color key beside the matrix (e.g. red denotes single genes).}
 \label{fig:Supp10}
\end{figure*}

\end{document}